\documentclass{jps-cp}
\usepackage{txfonts} 

\usepackage{amsmath,amscd,amssymb}
\usepackage{color,epsfig}
\usepackage{xfrac}
\usepackage{latexsym}
\usepackage{mathrsfs}
\usepackage{amsfonts}
\usepackage{mathtools}
\usepackage{xfrac}
\usepackage{slashed}
\usepackage{cite}

\newcommand{\imu}{{\rm i}}

\title{Structure Functions of the Nucleon in a Soliton Model}

\author{Ishmael \textsc{Takyi} and Herbert \textsc{Weigel}}

\inst{Institute of Theoretical Physics, Physics Department, 
Stellenbosch University, Matieland 7602, South Africa}

\email{takyi@sun.ac.za and weigel@sun.ac.za}

\abst{We study nucleon structure functions in the soliton picture of the 
bosonized Nambu-Jona-Lasinio model. We focus on their vacuum contributions
and examine whether they are outweighed by their valence quark counterparts.}

\kword{Nambu-Jona-Lasinio model, Chiral soliton, Nucleon structure functions}

\begin{document}
\maketitle

\section{Introduction}
We study nucleon structure functions within the Nambu-Jona-Lasinio (NJL) model. In this model
the nucleon emerges as a chiral soliton that polarizes the quark fields. Stability of the soliton
is achieved by balancing the binding energies of the valence levels 
and the vacuum polarization energy~\cite{Alkofer:1994ph}. 
The Compton tensor is a nucleon matrix element of a time ordered product and can be computed 
from the regularized action. Since the hadron tensor is the absorptive part of the Compton tensor this 
is the ideal point of departure to unambiguously extract the contributions of the polarized vacuum to 
the structure functions. These contributions have never before been computed directly from the fully 
regularized\cite{ft1} action and we numerically examine whether or not they considerably modify their valence 
counterparts. We also investigate the sum rules entering these structure functions. In addition to
their physical content they serve as consistency checks on the heavy numerical endeavor.
Finally the total structure functions undergo a perturbative DGLAP evolution to enable the
comparison with experimental data.

\section{The NJL Chiral Soliton Model}
The simplest $SU(2)$ NJL model only contains scalar and pseudoscalar fields. Its Lagrangian
is given by~\cite{Nambu:1961tp}
\begin{equation}
\mathcal{L}_{NJL}(q) = \bar{q} \left( \imu \slashed{\partial} -m^{0} \right) q + 
\frac{G}{2}\left[\left(\bar{q}q\right)^{2}+\left(\bar{q} \imu \gamma_{5}\vec{\tau\,}q\right)^{2} \right]\,,
\label{e1} \end{equation}
where $G$ is the coupling constant of the chirally invariant four fermion interaction.
Introducing scalar ($S$) and pseudoscalar ($P$) meson $2\times2$ matrix fields that 
couple to $\bar{q}q$ and $\bar{q} \imu \gamma_{5}\vec{\tau\,}q$, respectively, as $M=S+\imu P$
allows to integrate out the quark fields. This yields the effective action 
\begin{align}
\mathcal{A}[M,M^{\dagger}] &= -\frac{1}{4G} \int \mathrm{d}^{4} x\, 
\mathrm{tr} \left[M M^{\dagger} - m^{0} (M + M^{\dagger}) \right]  
- \imu N_{C} \mathrm{Tr}_{\Lambda} \log \Big\{ 
\imu \slashed{\partial} - \left( MP_{R} +  M^{\dagger}P_{L} \right) \Big\}\,. 
\label{e2} \end{align}
This action is quadratically divergent requiring regularization. For definiteness we adopt 
the Pauli-Villars subtraction scheme with a single cut-off $\Lambda$.
In this model chiral symmetry is dynamically broken as reflected by the non-zero
vacuum expectation value, $\langle S \rangle = m$.
In total there are three model parameters, $\Lambda$, $G$ and the current quark 
mass $m^{0}$. We identify the fluctuations $P$ as the pion field and impose
the empirical values $m_{\pi}=\mathrm{138\, MeV}$ and $f_{\pi} = \mathrm{93\, MeV}$ so that
$m$ is the only tunable parameter.

To construct the static soliton configuration, we impose the hedgehog ansatz 
which defines the Dirac Hamiltonian $h=-\imu \vec{\alpha}\cdot\vec{\partial}
+\beta m \exp \left(\imu\vec{\tau}\cdot\hat{r}\gamma_{5}F(r)\right)$. Its diagonalization
produces the energy eigenvalues $\epsilon_{\alpha}$ and the eigenspinors $\Psi_\alpha(\vec{r})$
as functionals of the chiral angle $F(r)$. Computing the functional trace in Eq.~\eqref{e2} 
in that basis yields the classical energy functional
\begin{align}
E[F(r)]  = \frac{m_\pi^2 f_\pi^2}{4}\int d^3r \left[1-\cos(F(r))\right]+
N_c\Bigg[ \eta_{\mathrm{v}}\epsilon_{\mathrm{v}}  
-\frac{1}{2}\sum_{\alpha} \Big\{|\epsilon_{\alpha}|-\sqrt{\epsilon_{\alpha}^2+\Lambda^2}
+\frac{\Lambda^2}{2\sqrt{\epsilon_{\alpha}^2+\Lambda^2}}\Big\}\Bigg] \,,
\label{e3}
\end{align}
the subscripts `$\mathrm{v}$' denotes the distinct valence level which is strongly bound 
in the soliton background and is added to guarantee unit baryon number via its
occupation number $\eta_{\mathrm{v}}=[1+\mathrm{sign}(\epsilon_{\mathrm{v}})]/2$. 
The chiral angle $F(r)$ is obtained by extremizing $E[F(r)]$ \eqref{e3}\cite{Alkofer:1994ph}.
Finally, nucleon states are generated by canonically quantizing the zero modes
of the soliton. Then the nucleon wave function depends on the $SU(2)$ rotation matrix $A$
that parameterizes the flavor rotational zero modes.

\section{Hadronic Tensor in the NJL Model}
Deep inelastic scattering measures the hadronic tensor which is obtained from the 
nucleon matrix element of the commutator of two electromagnetic currents 
$\left[ J_{\mu}(\xi), J_{\nu}(0)\right]$. In the Bjorken limit, when the
four-momentum $q$ of the virtual photon approaches negative spacelike infinity, the form 
factors of this tensor turn into structure functions that only depend on the 
Bjorken variable $x=-q^2/2q\cdot p$, where $p$ is the nucleon momentum. In the NJL 
model the electromagnetic current is written as $J_{\mu}=\bar{\psi}\mathcal{Q}\gamma_{\mu} \psi$, 
where $\mathcal{Q}$ is the flavor quark charge matrix. We wish to compute expectation 
values of products of these currents by introducing an auxiliary source field in the bosonized action and 
taking appropriate derivatives. The bosonized action, Eq.~\eqref{e2} is obtained from a 
path integral formalism in which such derivatives yield (matrix elements of) time ordered 
products. In case of two current operators, this is the Compton tensor. We then take advantage 
of the fact that the hadronic tensor equals the absorptive part of the Compton tensor.
Details of this analysis are reported in Ref. \cite{Weigel:1999pc}. For orientation we display
the resulting leading term in the $1/N_C$ counting for the vacuum contribution to the hardonic 
tensor
\begin{align}
W_{\mu\nu}(q) & = -\imu \frac{ M N_{C} \pi}{8} \int \frac{\mathrm{d}\omega}{2\pi}  
\sum_{\alpha} \int \mathrm{d}^{3} \xi \int \frac{\mathrm{d}\lambda}{2 \pi}\, e^{\imu Mx \lambda} 
\label{e6}\\
& \times \langle N,\vec{s} \hspace{0.1cm} | \Bigl\{ \Bigl[ \bar{\Psi}_{\alpha}(\vec{\xi}) 
\mathcal{Q}_{A}^{2} \gamma_{\mu}\slashed{n}\gamma_{\nu}\Psi_{\alpha}(\vec{\xi}+\lambda \hat{q}) 
e^{-\imath \lambda \omega} 
- \bar{\Psi}_{\alpha} (\vec{\xi}) \mathcal{Q}_{A}^{2} \gamma_{\nu} \slashed{n} \gamma_{\mu} 
\Psi_{\alpha} (\vec{\xi} - \lambda\hat{q}) e^{\imu \lambda \omega} \Bigr] \left. 
f_{\alpha}^{+} (\omega)\right|_{\text{p}} \cr
&  +\Bigl[ \bar{\Psi}_{\alpha}(\vec{\xi}) \mathcal{Q}_{A}^{2} 
\left( \gamma_{\mu}\slashed{n}\gamma_{\nu}\right)_{5}\Psi_{\alpha}(\vec{\xi}-\lambda\hat{q}) 
e^{-\imu \lambda \omega} 
- \bar{\Psi}_{\alpha} (\vec{\xi} ) \mathcal{Q}_{A}^{2}  \left( \gamma_{\nu} \slashed{n} \gamma_{\mu} \right)_{5} 
\Psi_{\alpha} (\vec{\xi} + \lambda \hat{q}) e^{\imath \lambda \omega}\Bigr] \left. f_{\alpha}^{-} (\omega)\right|_{\text{p}}
\Bigr\} | N,\vec{s} \hspace{0.1cm} \rangle\,,
\nonumber \end{align}
where $n^{\mu}=\left(1, \hat{q} \right)^{\mu}$ is the light-cone vector defined by the 
direction of the virtual photon. In the above
\begin{equation}
f_{\alpha}^{\pm}(\omega)=\frac{\omega \pm \epsilon_{\alpha}}{\omega^{2}-\epsilon_{\alpha}^{2}+\imu\epsilon}
-\frac{\omega \pm \epsilon_{\alpha}}{\omega^{2}-\epsilon_{\alpha}^{2}-\Lambda^2+\imu\epsilon}
+\Lambda^{2}\frac{\omega\pm\epsilon_{\alpha}}
{\left(\omega^{2}-\epsilon_{\alpha}^{2}-\Lambda^2+\imu\epsilon\right)^2}
 \pm \frac{\omega \pm \epsilon_{\alpha}}{\omega^{2} - \epsilon_{\alpha}^{2} + \imu \epsilon}, 
\label{e7} \end{equation}
are Pauli-Villars regularized spectral functions with 'p' extracting their
pole contributions. Furthermore $\mathcal{Q}_{A}=A\mathcal{Q}A^\dagger$ is the flavor rotated quark 
charge matrix. The subscript '5' refers to a particular treatment of the axial component of
$\gamma_{\mu}\slashed{n}\gamma_{\nu}$ for consistency of regularization~\cite{Weigel:1999pc,ft2}. 
It now remains to adopt particular components and kinematics to project Eq.~\eqref{e6} onto the relevant structure functions.
 
\section{Numerical Results}
In these proceedings we can only present a small sample of the vast numerical results for 
nucleon structure functions in the NJL model. More details will be presented elsewhere \cite{Takyi2019}.

Once the structure functions are computed from Eqs.~\eqref{e6} and~\eqref{e7} the valence quark counterparts need to be added. They were computed earlier and assumed to be dominant \cite{Weigel:1996kw,Weigel:1996jh,ft3}. 
The numerical simulation is quite costly because reliable Fourier transforms of all eigenfunctions 
$\Psi_\alpha$ must be obtained. Whenever applicable we have verified sum rules that relate integrated 
structure functions to coordinate space matrix elements of the $\Psi_\alpha$.

\subsection{Unpolarized Structure Functions}
\begin{figure}[t]
\centerline{
\includegraphics[width=9cm,height=3.1cm]{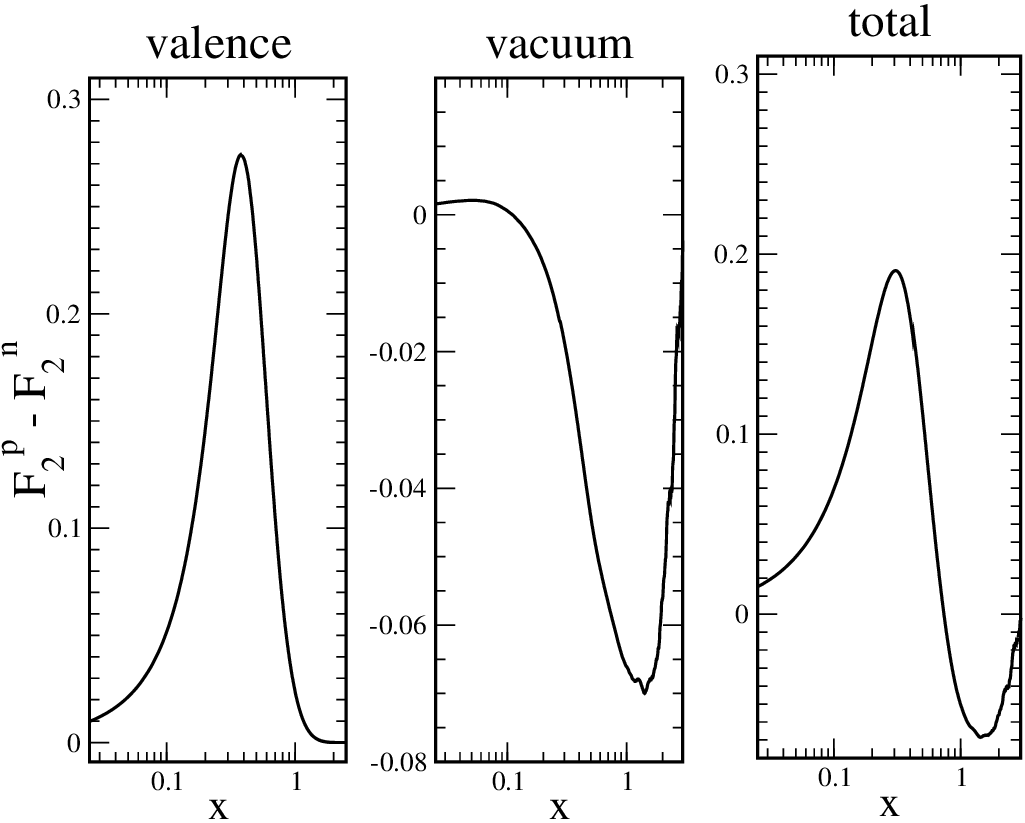}\hspace{1cm}
\includegraphics[width=5cm,height=3cm]{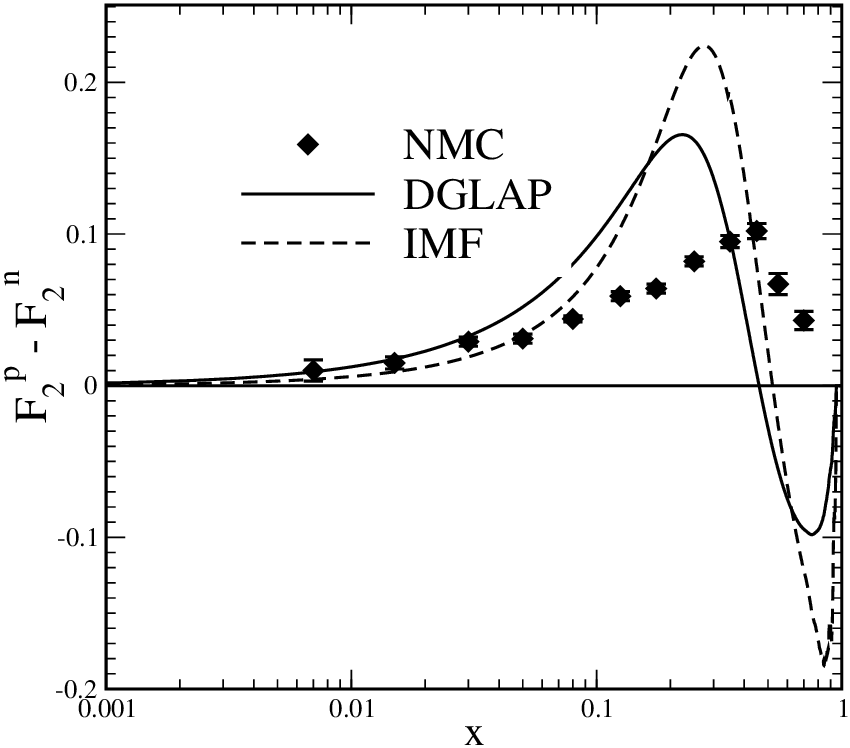}
}
\caption{Model prediction for the unpolarized structure function that enters the 
Gottfried sum rule. Left panel: valence and vacuum decomposition (note the different
scales); right panel: comparison of 'total' to empirical data \cite{Arneodo:1994sh} after projection (IMF) 
and evolution (DGLAP).}
\label{f1}
\end{figure}

In figure \ref{f1} we show the numerical results of the unpolarized structure 
function that enters the Gottfried sum rule using $m=\mathrm{400\, MeV}$. 
Indeed we see that the valence part dominates. Since the soliton breaks translational
invariance, the model structure functions are not guaranteed to be localized in $x \in \left[0,1\right]$.
This is particularly reflected by the vacuum part having a small negative contribution slightly above $x=1$.
Translational invariance is restored by transformation to the infinite momentum frame (IMF) 
\cite{Gamberg:1997qk}. Then the structure functions vanish for $x>1$
and are subjected to perturbative QCD evolution (DGLAP formalism). The resulting structure function
is compared to data \cite{Arneodo:1994sh} in the right panel of figure \ref{f1}. Though the gross 
structure is reproduced, in the vicinity of $x\lesssim1$ the negative part of the vacuum contribution
has an inauspicious impact. The integral 
$\mathcal{S}_{G}=\int_{0}^{\infty}\frac{\mathrm{d}x}{x}\,\left(F_{2}^{p}-F_{2}^{n}\right) $ is the 
Gottfried sum rule and we list our model prediction in table \ref{t1}. As there are obvious cancellations
when integrating the vacuum part, the total sum rule essentially equals its valence contribution.
\begin{table}[b]
\caption{\label{t1}The Gottfried sum rule for various values of $m$. The subscripts 'v' and 's'
denote the valence and vacuum contributions, respectively. The last column contains their sums.}
\begin{center}
\begin{tabular}{|c| c| c| c|}
\hline 
$m\,[\mathrm{MeV}]$  & $[\mathcal{S}_{G}]_{\mathrm{v}}$  & $[\mathcal{S}_{G}]_{\mathrm{s}}$  & $\mathcal{S}_{G}$\\
\hline
$400$  &  $0.214$   & $0.000156$  &  $0.214$   \\
$450$  &  $0.225$   & $0.000248$  &  $0.225$ \\
$500$  &  $0.236$   & $0.000356$  &  $0.237$ \\
\hline 
\end{tabular}
\end{center}
\end{table}
In total we obtain reasonable agreement with the experimental value $0.235\pm 0.026$\cite{Arneodo:1994sh}; 
in particular when confronted with the naive parton model prediction of 1/3.

\subsection{Polarized Structure Functions}
\begin{figure}[t]
\centerline{
\includegraphics[width=7cm,height=3cm]{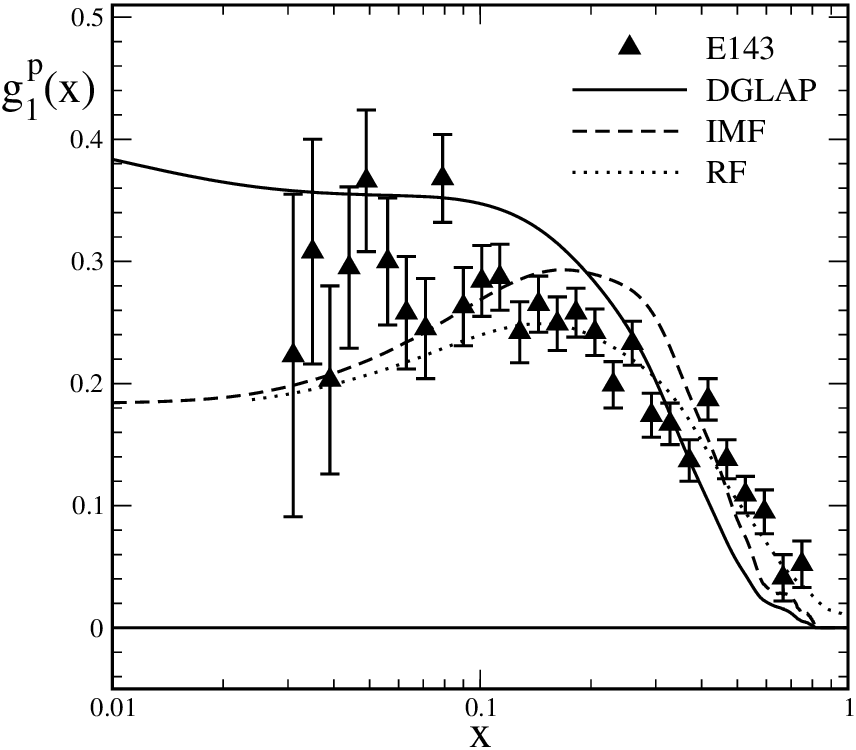}\hspace{1cm}
\includegraphics[width=7cm,height=3cm]{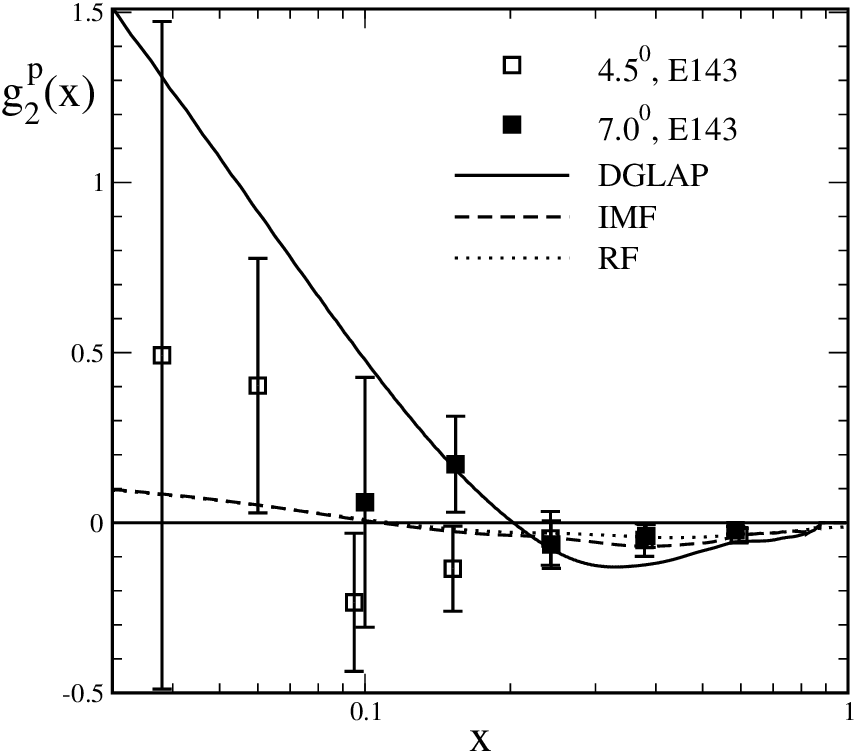}}
\caption{\label{f2}Predicted polarized structure functions computed for $m=400\mathrm{MeV}$.
The entry 'RF' refers to the actual model calculation while IMF and DGLAP denote projection
and evolution, respectively. Data are from Refs.~\cite{Abe:1994cp,Abe:1995dc}.}
\end{figure}

The polarized spin structure functions $g_{1}(x)$ and $g_{2}(x)$ are obtained from the antisymmetric 
contribution $W_{\mu \nu}(q) -W_{\nu\mu}(q)$. In figure \ref{f2} we show typical results for the axial 
structure functions of the proton. The data are well produced. When combined with the neutron, the 
corresponding (so-called Bjorken) sum rule gives the axial charge $g_A$ whose empirical value is 
$1.2601\pm0.0025$\cite{Barnett:1996hr}. As typical in soliton models, this value is 
underestimated by about 30\%-40\%, {\it cf.} table \ref{t2}. Yet the computed axial 
singlet charge, which is subleading in $1/N_C$, agrees with the empirical value
$\Delta \Sigma \sim 0.31 \pm 0.07$ \cite{Ellis:1994py}. 

\begin{table}[t]
\vskip-0.3cm
\caption{\label{t2}The axial-vector and -singlet charges for various values of $m$. 
Subscripts are as in table \ref{t1}.}
\begin{center}
\begin{minipage}[b]{0.05\linewidth}~~~~\end{minipage}
\begin{minipage}[b]{0.40\linewidth}
\begin{tabular}{|c| c| c| c|}
\hline 
$m\,[\mathrm{MeV}]$  & $[ g_{A}]_{\mathrm{v}}$ & $ [ g_{A}]_{\mathrm{s}}$  & $ g_{A}$  \\
\hline
$400$  &  $0.734$  & $0.0648$ &  $0.799$     \\
$450$  &  $0.715$  & $0.0509$ &  $0.766$    \\
$500$  &  $0.704$  & $0.0289$ &  $0.733$ \\
\hline 
\end{tabular}
\end{minipage}
\quad
\begin{minipage}[b]{0.40\linewidth}
\begin{tabular}{|c| c| c| c|}
\hline 
$m\,[\mathrm{MeV}]$  & $[ g_{A}^{0}]_{\mathrm{v}}$ & $[ g_{A}^{0}]_{\mathrm{s}}$ & $ g_{A}^{0}$  \\
\hline
$400$  &  $0.344$  & $0.00157$  &  $0.345$   \\
$450$  &  $0.327$  & $0.00214$  &  $0.329$    \\
$500$  &  $0.316$  & $0.00282$  &  $0.318$    \\
\hline 
\end{tabular}
\end{minipage}
\end{center}
\vskip-0.2cm
\end{table}

\section{Conclusion}
We have computed nucleon structure functions from the chiral soliton of the
bosonized NJL model. This approach has the very important feature that the 
regularization of the vacuum contribution is self-contained when computed 
from the Compton tensor. Our numerical simulations
confirm that indeed
the valence level contribution to the structure functions dominates over the 
vacuum counterpart. In comparison with data we find some short-comings for the 
unpolarized structure functions while the polarized structure functions of the 
proton are nicely reproduced.

\section{Acknowledgment}
One of us (IT) would like to thank the organizers for making QNP 2018 a worthwhile 
conference. This work was supported in part by the Stellenbosch Institute for Advanced 
Studies (STIAS) and by funds provided by the National Research Foundation of South Africa
(NRF) under contract~109497.


\begin{thebibliography}{9}
\bibitem{Alkofer:1994ph} R. Alkofer, H. Reinhardt, and H. Weigel, Phys. Rep. \textbf{265}, 139 (1996).
\bibitem{ft1} Computations with {\it ad hoc} regularization procedures were reported in \\
D.~Diakonov et al.,
  Phys.\ Rev.\ D {\bf 56}, 4069 (1997); Nucl.\ Phys.\ B {\bf 480}, 341 (1996);\\
M.~Wakamatsu and T.~Kubota,
  Phys.\ Rev.\ D {\bf 57}, 5755 (1998); Phys.\ Rev.\ D {\bf 60}, 034020 (1999).
\bibitem{Nambu:1961tp} Y. Nambu and Jona-Lasinio, Phys. Rev. \textbf{122}, 345 (1961).
\bibitem{Weigel:1999pc} H. Weigel, E. Ruiz-Arriola and L. Gamberg, Nucl. Phys. \textbf{B560}, 383 (1999).
\bibitem{ft2}Note that $f_\alpha^{-}(\omega)=0$ without regularization.
\bibitem{Takyi2019} I. Takyi and H. Weigel, forthcoming.
\bibitem{Weigel:1996kw} H. Weigel, L. Gamberg, H. Reinhardt, Phys. Lett. \textbf{B399}, 287 (1997).
\bibitem{Weigel:1996jh} H. Weigel, L. Gamberg, H. Reinhardt, Phys. Rev. \textbf{D55}, 6910 (1997).
\bibitem{ft3}They are also obtained from Eq.~\eqref{e6} by setting 
$f_\alpha^{+}(\omega)=2 \eta_v\delta_{\alpha v}\delta(\omega-\epsilon_v)$ 
and $f_\alpha^{-}(\omega)=0$.
\bibitem{Gamberg:1997qk} L. Gamberg, H. Reinhardt, and H. Weigel, Int. J. Mod. Phys. \textbf{A13}, 5519 (1998).
\bibitem{Arneodo:1994sh} M. Arneodo et al., Phys. Rev. \textbf{D 50}, R1 (1994).
\bibitem{Barnett:1996hr} R. M. Barnett et al., Phys. Rev. \textbf{D54}, 1 (1996).
\bibitem{Abe:1994cp} K. Abe et al., Phys. Rev. Lett. \textbf{74}, 346 (1995).
\bibitem{Abe:1995dc} K. Abe et al., Phys. Rev. Lett. \textbf{76}, 587 (1996).
\bibitem{Ellis:1994py} J. Ellis and M. Karliner, Phys. Lett. \textbf{B341}, 397 (1995).
\end{thebibliography}
\end{document}